# A method to suppress polar Kerr signal in a longitudinal magneto-optic-Kerr-effect measurement

Ryan W. Greening*, Elyssa D. DeVisscher*, Xin Fan

*Department of Physics and Astronomy, University of Denver, Denver, CO 80210, USA*

* These authors contribute equally.

Abstract:

The Magneto-Optical-Kerr-Effect (MOKE) is a convenient technique to study the magnetization of thin films. However, both polar and longitudinal MOKE responses contribute to the total Kerr response in a typical longitudinal MOKE measurement. Here, we present a simple optical technique to suppress the polar MOKE response in the oblique angle incidence by exploiting differences between polar and longitudinal MOKE responses upon double reflection from the sample. By using a mirror to reflect the beam and by selectively using a quarter waveplate, the polar or longitudinal MOKE signals can be suppressed, and therefore studied separately using the same oblique experimental setup. To demonstrate the feasibility of this technique, we use an out-of-plane magnetized Pt/Co/Pt film as well as a Pt/Co/Cu/NiFe Heterostructure with both in-plane and out-of-plane magnetization. We show that the polar MOKE of the CoPt film can be suppressed by a factor of 6 compared to a conventional MOKE measurement. By accounting for birefringence, we further reduce the polar MOKE response in a longitudinal MOKE measurement of the Pt/Co/Cu/NiFe film by over 160 times compared to a conventional oblique-angle MOKE measurement.

The Magneto-Optical-Kerr-Effect (MOKE), discovered in 1877 by John Kerr [Ref. 1, 2], is widely used to study magnetization hysteresis, magnetic textures, magnetization dynamics, and spin-orbit-torques in magnetic thin films [Ref. 3-15]. The MOKE describes the rotation of the polarization of light upon reflection from a magnetic film [Ref. 2, 16].

The so-called Kerr rotation can be due to either the polar or longitudinal MOKE, which corresponds to the out-of-plane magnetization and the in-plane magnetization in the incidence plane, respectively [Ref. 17]. If the incidence beam is normal to the film surface, the longitudinal MOKE is suppressed, and the MOKE response is due solely to the polar MOKE. To measure the in-plane magnetization, an oblique angle incidence is needed. However, in this oblique-angle configuration, the measured signal has contributions from both the polar and longitudinal MOKE responses [Ref. 11, 18]. Since polar MOKE is typically an order of magnitude larger than the longitudinal MOKE, the extraneous polar MOKE signal must be accounted for in the combined signal [Ref. 3].

The polar and longitudinal MOKE have different dependencies on the direction of the incidence beam, because the rotational symmetry about the Z-axis is broken by the in-plane magnetization, but sustained by the out-of-plane magnetization [Ref. 19]. As illustrated in Fig. 1(a), when the incidence beam reverses direction, the Kerr rotation for the polar MOKE is unchanged, while that of the longitudinal MOKE is reversed, due to the different symmetries of each case.

Based on the differential symmetries, Ding *et al.* [Ref. 19] suggested separation of the polar and longitudinal MOKE by measuring the MOKE response, then swapping the position of the laser source and detector and remeasuring in the new configuration. Taking advantage of the different dependence on incident angle with respect to the magnetization of the film, the two measurement results are added and subtracted to extrapolate the polar and longitudinal MOKE contributions independently. However, this technique requires significant rearrangement of optical components and their realignment. Inconsistencies in the alignment may also lead to errors in the extrapolation of longitudinal and polar MOKE signals. An additional technique developed involves quadrant-based detection with a normal-incidence light [Ref. 20-23]. In one such experiment performed by Celik *et al.*, the light is tightly focused onto the sample using an objective lens, producing some light components with oblique angle incidence. The reflected beam is analyzed with a quadrant detector, and by taking sums and differences from the signals in each of the 4 quadrants, they find the contributions from the polar, longitudinal, and quadratic responses, without altering the experimental setup for each response [Ref. 20].

Here we propose an alternative experimental setup to suppress the polar MOKE signal in the longitudinal MOKE measurement, which uses a mirror to reflect the beam back onto the sample. To suppress the polar MOKE signal, a quarter-wave plate is inserted into the optical path. By simply removing and replacing this quarter-wave plate, switching between the polar- and longitudinal-MOKE-sensitive configurations is easily achieved. Reflecting the beam back onto the sample can take advantage of the rotational symmetry about the Z-axis, which will in principle cancel out either the polar or longitudinal MOKE signal as shown in Fig. 1(a).

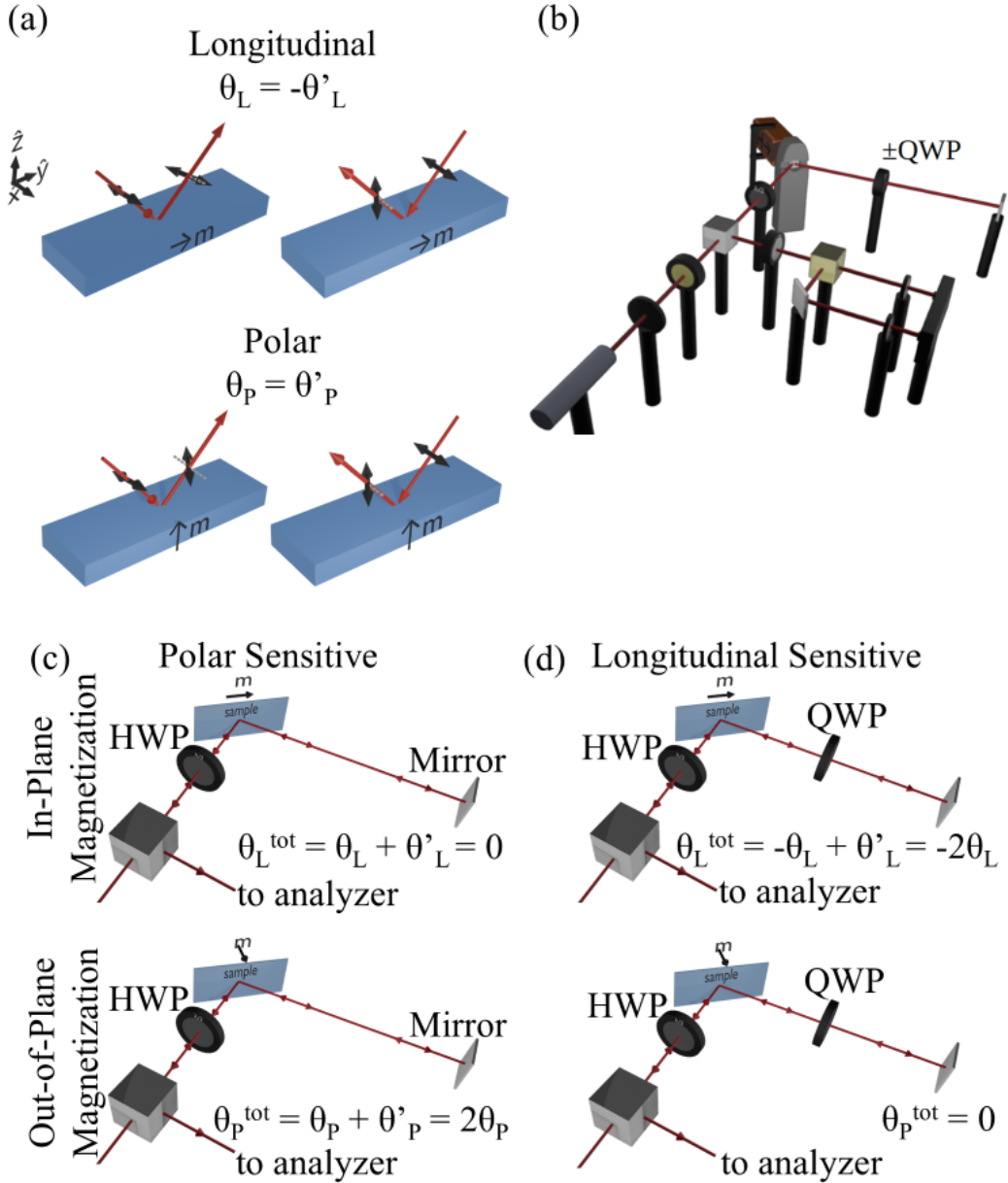

Figure 1. Schematic of the experimental setup. Fig. 1(a). The returning incidence beam reverses the Kerr rotation in the longitudinal case, and maintains the Kerr rotation in the polar case. The gray dotted line indicates the unrotated polarization. Fig. 1(b). The optics that comprise the longitudinal-MOKE-sensitive setup (including the QWP), and the optics that comprise the polar-MOKE-sensitive setup (excluding the QWP). Fig. 1(c). The polar-MOKE-sensitive configuration suppresses the longitudinal signal of a sample with in-plane magnetization and enhances the polar signal of a sample with out-of-plane magnetization. Fig. 1(d). The longitudinal-MOKE-sensitive configuration enhances the longitudinal signal of a sample with in-plane magnetization and suppresses the polar signal of a sample with out-of-plane magnetization.

In the polar MOKE case, rotational symmetry is retained ($\theta_P = \theta_P'$), and the longitudinal signal is suppressed. In the longitudinal MOKE case, rotational symmetry is broken ($\theta_L = -\theta_L'$) [Ref. 19], and the polar signal is suppressed by a quarter-wave plate. Because the beam passes through it twice, the quarter-wave plate effectively acts as a half-wave plate. As a result, the polarization is flipped about the fast axis, and the Kerr rotation of the reflected light is reversed. The returning light beam is then reflected by the sample a second time, undergoing another Kerr rotation.

Depending on whether the quarter-wave plate is inserted into the optical path, the longitudinal or polar Kerr rotation is selectively enhanced or suppressed, as shown in Fig. 1(c, d). The configuration with the quarter-wave plate inserted into the optical path is the longitudinal-sensitive configuration, whereas the configuration without the quarter-wave plate is referred to as the polar-sensitive configuration (Fig. 1(c, d)).

In this measurement, the sample is mounted in the center of an electromagnet, which can produce both in-plane and out-of-plane magnetic fields. A Coherent ultra low noise diode laser with 5 mW nominal power and 635 nm wavelength is used as the laser source. As shown in Fig. 1(b), the laser passes through a neutral density filter and a polarizer so that the light is p-polarized with respect to the plane of incidence. The laser then passes through a 50/50 power beam splitter and half-wave plate mounted in a rotation mount. The light is incident to the sample at approximately 45°.

In the polar-MOKE-sensitive configuration, the reflected beam is reversed by the mirror, and the beam is reflected onto the sample a second time. The beam reflected by the sample is steered by the reflection mirror onto an iris to ensure the original incident and reflected beams are coincident. Following reflection from the sample, the beam is directed by the 50/50 power beam splitter into the analyzer. The analyzer, which analyzes the polarization change, consists of a rotatable half-wave plate, a polarization beam splitter, and a balanced beam detector. The polarizing beam splitter separates the beam into s- and p-polarized components, allowing each polarization component to be focused onto the diodes of the balanced beam detector. The half-wave plate is used to balance the two beams' intensity after the polarization beam splitter. Using a balanced beam detector, the resultant signal from each photodiode is subtracted to obtain a power difference between the s- and p-polarized light. The voltage difference is then normalized by blocking one side of the balanced beam detector to determine the voltage on each arm, which can be used to determine the angle of the Kerr rotation in milliradians.

The only difference in the longitudinal-sensitive configuration is that a quarter-wave plate is inserted in between the reflection mirror and the sample. The principal axis of the quarter-wave plate is aligned with its fast axis along the polarization of the incident light.

To measure Kerr rotation under conventional oblique angle incidence, a linear polarizer is placed between the film and the reflection mirror. The polarizer is set so that its principal axis is aligned with the initial polarization of the beam. Any polarization rotation or ellipticity induced by the film or quarter-wave plates in the longitudinal-sensitive configuration will be eliminated by the polarizer and the reflected beam will be in the original polarization configuration, but reduced in intensity. This linearly polarized beam will then reflect off of the sample and be analyzed. This setup is equivalent to performing conventional oblique-angle MOKE without requiring major modifications to the experimental apparatus.

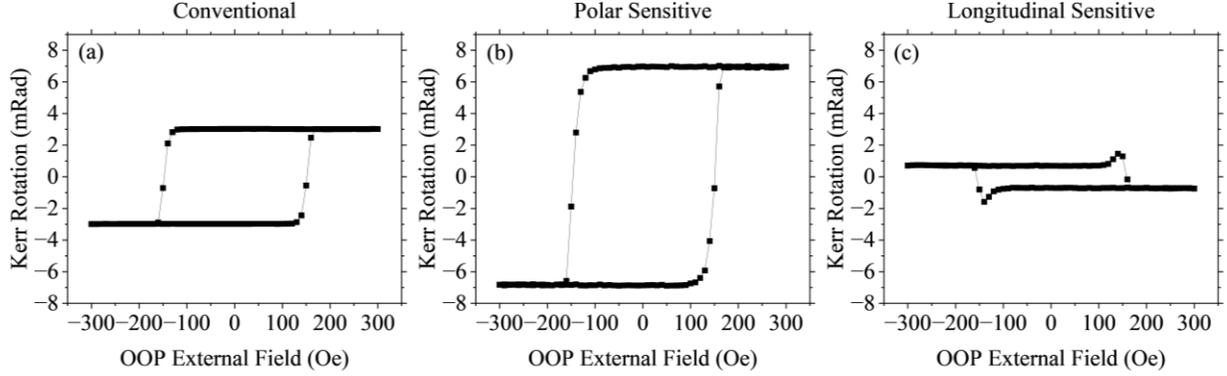

Figure 2. Kerr rotation for Ta/Pt/Co/Pt film with three configurations. Fig. 2(a). Out-of-plane sweeping in a conventional longitudinal MOKE configuration. Fig. 2(b). Out-of-plane sweeping in the polar-sensitive configuration Fig. 2(c). Out-of-plane sweeping in the longitudinal-sensitive configuration.

To test polar MOKE suppression, a Pt(3)/Co(0.5)/Pt(3)/Ta(3)//Sub film is used, where numbers in parentheses indicate layer thickness in nanometers. This film is perpendicularly magnetized due to the perpendicular anisotropy induced by the Ta/Pt seed layer. The film is mounted, and the hysteresis is measured under an out-of-plane external field sweep following the experimental technique described above.

As a baseline, the out-of-plane magnetization hysteresis is first measured in the conventional configuration, as shown in Fig. 2(a), yielding a peak-to-peak difference of $5.99 \pm 0.01$ mrad. The polar-MOKE-sensitive configuration (Fig. 1(c)) is then employed to measure the out-of-plane magnetization hysteresis again, resulting in a clear hysteresis with a nearly doubled peak-to-peak difference of $13.81 \pm 0.01$ mrad (Fig. 2(b)). We then used the longitudinal-sensitive configuration (Fig. 1(d)) by inserting a quarter-wave plate. As illustrated in Fig. 2(c), the square hysteresis due to the polar MOKE signal is suppressed to a peak-to-peak difference of $1.39 \pm 0.01$ mrad, which is a suppression of the signal by nearly 6 times compared to the conventional MOKE.

Moreover, two distinct peaks near the coercivity are observed, coinciding with out-of-plane magnetization switching. These features may be attributed to minor misalignment of the incident and reflected beams, which may sample different domains during out-of-plane switching and lead to incomplete cancellation.

The signals in Fig. 2(c) exhibit significant but not complete suppression. The incomplete suppression is attributed to the birefringence of the measured films. When p-polarized light is incident on the sample, the resulting Kerr rotation can be decomposed into s and p components, with the Kerr rotation lying along the s component. Because birefringent films reflect differently depending on the incident polarization, the beam returns with polarization in a superposition of s and p states, each undergoing Kerr rotation again with different reflection coefficients [Ref.13]. As a result, complete suppression of the polar MOKE signal is not achieved.

To analytically examine the experiment using the Jones matrix formalism [Ref. 24-27], a Jones matrix is employed for an arbitrary magnetic thin film with either in-plane or out-of-plane magnetization.

$$J[\pm] = \begin{bmatrix} A & qm_z \mp Qm_x \\ qm_z \pm Qm_x & B \end{bmatrix} \quad (1)$$

The factors A and B on the diagonals of the Jones matrix account for the different reflection coefficients due to birefringence. The off-diagonal terms correspond to the polar and longitudinal Kerr response, q and Q respectively, while $M_z$ and $M_x$ are the magnetization direction of the material. In general, in longitudinal MOKE Kerr rotation Jones matrices, the off-diagonal terms are equal yet opposite, while for polar MOKE they have the same sign [Ref. 24]. This is accounted for by the ± or ∓ signs in front of the longitudinal MOKE response in the Jones matrix. The ± or ∓ are further utilized to account for the impact of the direction of propagation. The plus and minus signs correspond to either the initial angle of incidence J[+] or the inverse angle of incidence after reflection from the mirror J[−]. The Jones matrix for our experiment can then be described by the following equation:

$$R[-\alpha].[HWP].R[\alpha].J[-].R[-\beta].[QWP].R[\beta].[HWP].R[\beta].[QWP].R[-\beta].J[+].R[\alpha].[HWP].R[-\alpha] \quad (2)$$

$$[HWP] = \begin{bmatrix} 1 & 0 \\ 0 & -1 \end{bmatrix}, \quad [QWP] = \begin{bmatrix} 1 & 0 \\ 0 & i \end{bmatrix}, \quad R[\theta] = \begin{bmatrix} \cos\theta & -\sin\theta \\ \sin\theta & \cos\theta \end{bmatrix}$$

Where the first three matrices describe how the half-wave plate rotates the polarization by an arbitrary angle α, the 4th term is the Jones matrix as described above. The terms in between the J[+] and J[−] MOKE terms describe how the polarization is rotated by the quarter-wave plate at an arbitrary angle β and is then reflected by a mirror. The reflection by the mirror induces a coordinate transformation which is accounted for by the addition of the half-wave plate matrix. The light then propagates back through the quarter-wave plate and interacts with the sample for a second time. This second MOKE Jones matrix is similar to the first, but the off-diagonal terms have a sign change to account for the impact of the opposite incidence angle compared to the initial incidence angle. Finally, the reflected light is rotated again by the original half-wave plate. By performing the matrix multiplication and neglecting higher order Q terms which correspond to quadratic MOKE, the relation is simplified to the following Jones matrix when the quarter-wave plate and half-wave plate are oriented with the fast axis at 0° with respect to the initial polarization state:

$$\begin{bmatrix} A^2 & (-A-B)qm_z + (A-B)Qm_x \\ (-A-B)qm_z + (A-B)Qm_x & B^2 \end{bmatrix} \quad (3)$$

Two cases are considered: (1) out-of-plane magnetization, with $m_z=1$ and $m_x=0$, and (2) in-plane magnetization, with $m_z=0$ and $m_x=1$. In these two cases, equation 3 can be reduced to:

$$\begin{cases} \begin{bmatrix} A^2 & (-A-B)qm_z \\ (-A-B)qm_z & B^2 \end{bmatrix} & OOP\ Mag. \\ \begin{bmatrix} A^2 & (A-B)Qm_x \\ (A-B)Qm_x & B^2 \end{bmatrix} & IP\ Mag. \end{cases} \quad (4)$$

If the off-diagonal elements of this matrix equal zero, the MOKE response to magnetization is inhibited. For the out-of-plane magnetized case, this can only occur if −A = B when birefringence is absent. Simultaneously for a film with in-plane magnetization, the Kerr rotation should be doubled compared to a conventional MOKE measurement. As evidenced in Fig. 2, complete suppression of the polar signal in the longitudinal MOKE configuration is not achieved due to birefringence.

It is hypothesized that tuning the incident polarization can offset the effect of birefringence to further suppress the polar MOKE response. To validate this, numerical simulations of magnetic thin films are performed. The Jones matrix for the MOKE response is computed using the propagation matrix method from [Ref. 24, 28], in conjunction with the Jones matrix for the experimental technique described above (Eq. 2). Once the MOKE Jones matrix is obtained for both forward and reverse propagation directions, the matrix multiplication is performed for each combination of half- and quarter-wave plate configurations. Fig. 3 presents the simulation results for an Air/Co(0.6)/Cu(3)/NiFe(3)/SiOx(1000) heterostructure on a Si substrate.

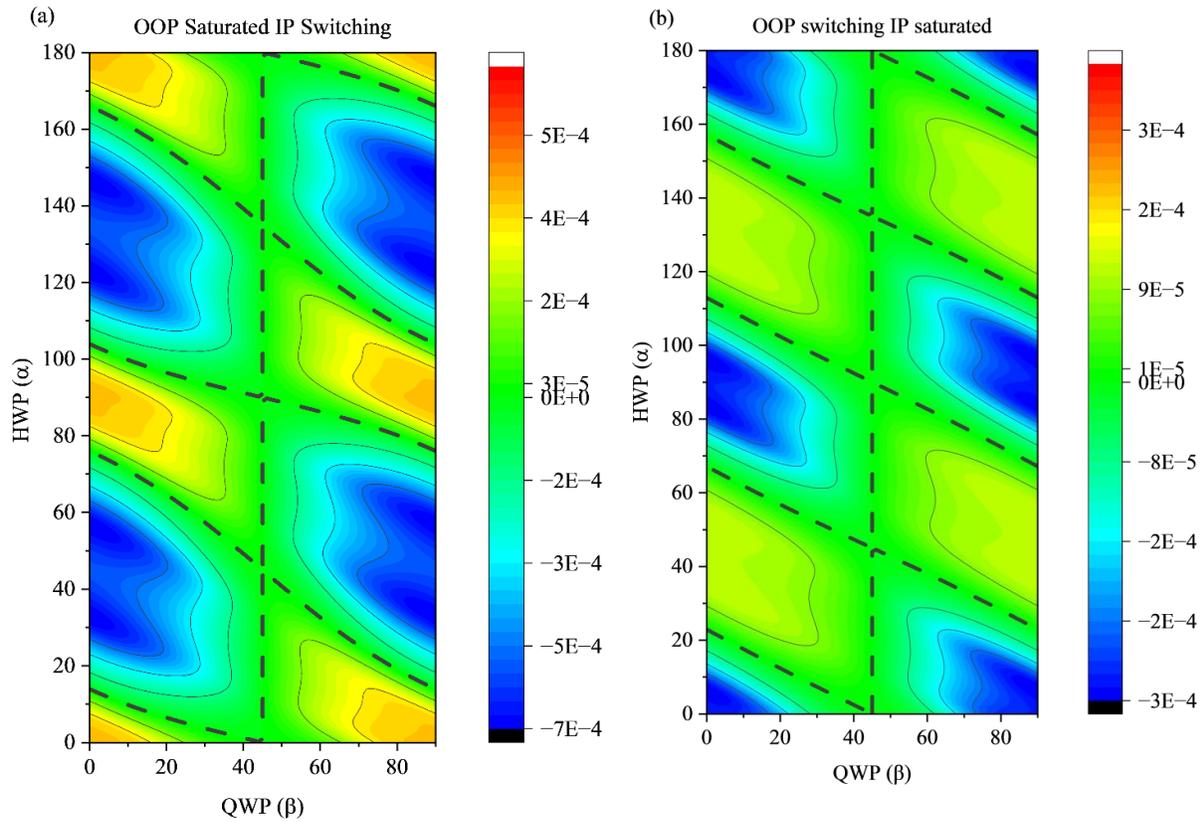

Figure 3. Numerical simulation results for two cases: Fig. 3(a). when the Co layer is saturated out-of-plane while the NiFe layer switches in-plane and Fig. 3(b). when the Co layer switches out-of-plane while the NiFe layer remains saturated in-plane.

For this experimental technique, any quarter- and half-wave plate configuration that yields zero polar Kerr rotation while maintaining non-zero longitudinal Kerr rotation is potentially valid. The dashed black lines in Fig. 3 represent the zero-isocline, indicating perfect suppression for those measurement conditions. Several potential solutions are observed when the quarter-wave plate is set to 0°, where the longitudinal-sensitive measurement suppresses the polar MOKE response but not the longitudinal MOKE signal. This indicates that, by varying the incident half-wave plate, a configuration can be achieved that provides complete polar suppression without eliminating the longitudinal MOKE response. Furthermore, when the quarter-wave plate is set to 45° relative to the incident polarization, both polar and longitudinal MOKE signals are universally suppressed, consistent with other observed results. These simulations suggest that empirically tuning the incident polarization allows for identifying half- and quarter-wave plate angles that account for birefringence effects.

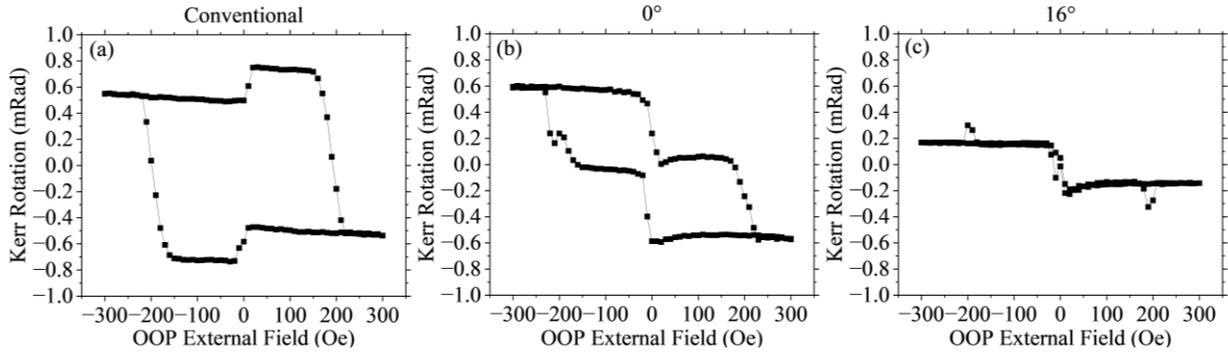

Figure 4. Kerr rotation for Co/Cu/NiFe trilayer with three configurations. Fig. 4(a). Out-of-plane sweeping in a conventional MOKE configuration. Fig. 4(b). Out-of-plane sweeping in the in-plane-sensitive configuration with the incident HWP tuned to 0°. Fig. 4(c). Out-of-plane sweeping in the in-plane-sensitive configuration with the incident HWP tuned to 16°.

To examine polar suppression in a sample that exhibits a large polar Kerr rotation signal, a trilayer structure with both in-plane and out-of-plane magnetized layers is fabricated. A stack of Ta(3)/Cu(3)/NiFe(3)/Cu(3)/Co(0.6)/Pt(3)/Ta(3)//Sub is grown. The NiFe layer is magnetized in-plane, while the Co layer is magnetized out-of-plane due to the perpendicular anisotropy induced by the Co/Pt interface. A 3 nm Cu layer is included to decouple the two magnetic layers. The sample is mounted, and the hysteresis is measured under a slightly tilted out-of-plane external field that induces both in-plane magnetization switching of the NiFe layer and out-of-plane magnetization switching of the Co layer.

In Fig. 4(a), the film's hysteresis under a conventional MOKE configuration is measured by placing a polarizer after the reflection mirror to cancel the Kerr rotation from the first sample reflection. In this configuration, the NiFe layer switching produces a longitudinal Kerr rotation of $0.26 \pm 0.00$ mrad, while the Co layer switching produces a polar Kerr rotation of $1.23 \pm 0.00$ mrad.

Using the in-plane magnetization–sensitive configuration described above, the quarter-wave plate is set to 0°, while the incident half-wave plate angle is varied from 0° to 45° to determine an angle under which the polar Kerr signal is completely suppressed. In Fig. 4(b), the film is measured with the half-wave plate set to 0° as a starting configuration. Under these conditions, a $0.49 \pm 0.00$ mrad signal from the NiFe layer and a $0.61 \pm 0.00$ mrad signal from the Co layer are observed. Compared to the conventional configuration in Fig. 4(a), the polar Kerr rotation signal is suppressed by only a factor of two, and the in-plane longitudinal signal from NiFe is enhanced by nearly the same amount.

To further suppress the polar Kerr rotation signal due to Co, the incident half-wave plate is adjusted to maximize polar Kerr rotation suppression, as predicted by the numerical simulations. A suitable configuration is identified with the half-wave plate set to 16°, as shown in Fig. 4(c). In this arrangement, a 0.33 ± 2.98E-3 mrad longitudinal signal corresponds to the NiFe in-plane layer switching, and a 7.56E-3 ± 2.98E-3 mrad polar signal corresponds to the perpendicular magnetized Co layer switching. Compared to the conventional MOKE configuration, this setting suppresses the polar MOKE Kerr rotation by a factor of more than 160 times.

Although the NiFe layer Kerr rotation at 16° is enhanced compared to the conventional MOKE configuration, it remains significantly smaller than the in-plane NiFe signal at 0°. This behavior is consistent with Fig. 3, which shows that the conditions for perfect polar MOKE suppression do not necessarily coincide with the conditions for maximal longitudinal MOKE signal. By tuning the half-wave plate, near-complete suppression of the polar MOKE Kerr rotation in the out-of-plane Co layer is achieved, while the longitudinal MOKE Kerr rotation of the NiFe layer is simultaneously enhanced.

In conclusion, a simple method is demonstrated to suppress polar MOKE signals in an oblique-angle incidence configuration by using a mirror and a quarter-wave plate. The polar MOKE signal is suppressed by more than 6 times compared to the conventional measurement configuration. By accounting for the impact of birefringence and empirically tuning a half-wave plate, the polar MOKE response under oblique incidence is fully suppressed while simultaneously enhancing the longitudinal signal relative to a conventional oblique-incidence MOKE measurement. By further adjusting the half-wave plate, a suppression factor exceeding 160 is achieved, although a factor of 60 is typically sufficient, given that the second-order (quadratic) MOKE effect is of a similar scale [Ref. 29]. This convenient technique is expected to improve the sensitivity and capabilities of MOKE-based magnetometry.

The authors have no conflicts of interest to disclose.

This research is supported by the National Science Foundation Award no. 2047118.

# Bibliography

16. Pershan, P. S. (1967). Magneto-optical effects. *Journal of Applied Physics, 38*, 1482–1490. https://doi.org/10.1063/1.1709678

17. Kerr, J. (1878). XXIV. On reflection of polarized light from the equatorial surface of a magnet. *The London, Edinburgh, and Dublin Philosophical Magazine and Journal of Science, 5*(30), 161–177. https://doi.org/10.1080/14786447808639407

18. Yang, Z. J., & Scheinfein, M. R. (1993). Combined three-axis surface magneto-optical Kerr effects in the study of surface and ultrathin-film magnetism. *Journal of Applied Physics, 74*(11), 6810–6823. https://doi.org/10.1063/1.355081

19. Ding, H., Putter, S., Oepen, H., & Kirschner, J. (2000). Experimental method for separating longitudinal and polar Kerr signals. *Journal of Magnetism and Magnetic Materials, 212*, 5. https://doi.org/10.1016/S0304-8853(99)00790-8

20. Celik, H., Kannan, H., Wang, T., Mellnik, A., Fan, X., Zhou, X., Barri, R., Ralph, D., Doty, M., Lorenz, V., Xiao, J. (2019). Vector-Resolved Magnetooptic Kerr Effect Measurements of Spin–Orbit Torque. *IEEE Transactions on Magnetics, 55*(1), 1–5, Art no. 4100105. https://doi.org/10.1109/TMAG.2018.2873129

21. Wright, C. D., Clegg, W. W., Boudjemline, A., & Heyes, N. A. E. (1994). Scanning Laser Microscopy of Magneto-Optic Storage Media. *Japanese Journal of Applied Physics, 33*, 2058. https://doi.org/10.1143/JJAP.33.2058

22. Hiebert, W. K., Ballentine, G. E., Lagae, L., Hunt, R. W., & Freeman, M. R. (2002). Ultrafast imaging of incoherent rotation magnetic switching with experimental and numerical micromagnetic dynamics. *Journal of Applied Physics, 92*(1), 392–396. https://doi.org/10.1063/1.1484225
12